\documentclass[twocolumn,showpacs,preprintnumbers,amsmath,amssymb]{revtex4}

\usepackage{graphicx}
\usepackage{dcolumn}
\usepackage{bm}

\begin{document}

\preprint{C2118/02/3.1}

\title{ On the occurrence of a phase transition in atomic systems }

\author{Luca Gamberale}
\email{Luca.Gamberale@pirelli.com}
\author{Daniele Garbelli}%
\email{Daniele.Garbelli@pirelli.com}
\affiliation{%
 Pirelli Labs - Materials Innovation\\
v.le Sarca 222, 20126 Milano (Italy)
}%
\begin{abstract}
We present a simple argument confirming the spontaneous quantum condensation of an electromagnetic field in matter in the case of a multi-level atomic system coupled to a single-mode electromagnetic field in the dipole approximation. The obtained result disproves the validity of a ``no-go theorem'' and fully rehabilitates theories of spontaneous quantum condensation of the electromagnetic field in matter. Peculiar features of the identified coherent states are outlined.
\end{abstract}

\maketitle

\section{\label{sec:level1}Introduction\protect}
The spontaneous appearance of a superradiant phase transition of the electromagnetic field in matter, first proposed by Dicke \cite{dicke} in 1954 and later by Hepp and Lieb \cite{hepplieb1}, Wang and Hioe \cite{wanghioe1} and Preparata \cite{prep1}, was questioned by Bialynicki-Birula and Rzazewski \cite{bbr},  who stated a ``no-go theorem'' on the condensation of a coherent em field in an ensemble of atoms in the dipole approximation.
This issue has been examined by Preparata {\it et al.} \cite{prepmele}, Enz \cite{enz} and Widom {\it et al.} \cite{widom}.  
Starting from a result obtained in Ref. \onlinecite{widom}, we show by explicit construction that at zero temperature and for a multi-level atomic system in the dipole approximation there exist coherent states whose energy is lower than that of any zero-field configuration when atomic density overcomes a critical value. The selected quantum state exhibits peculiar characteristics that allow for a possible explanation of coherent phenomena in condensed matter.
\section{\label{sec:level1.1} Dipole coupling of many atoms \protect}
A system of $N$
 atoms interacting with a single-mode electromagnetic field is described by the Hamiltonian \footnote{ Differently from Ref. \onlinecite{bbr}, we have included in $H_F$ both polarizations. This does not change the result. }
\begin{widetext}
\begin{subequations}
\label{eq:2.1} 
\begin{eqnarray}
 H& =& H_F  + H_A  \\
 H_F  &= &\hbar \omega \sum\limits_{s = 1,2} {c_s^\dag  c_s }  \\ 
 H_A  &=& \frac{1}{{2M}}\sum\limits_{i = 1}^N {\left( {\vec P_i  - \frac{{Ze}}{c}\vec A(\vec R_i )} \right)} ^2  + \frac{1}{{2m}}\sum\limits_{i = 1}^N {\sum\limits_{z = 1}^Z {\left( {\vec p_{iz}  + \frac{e}{c}\vec A(\vec r_{iz} )} \right)^2  + V\left( {\left\{ {\vec R_i } , {\vec r_{iz} } \right\}} \right) 
} } 
\end{eqnarray}
\end{subequations}
where $M,{\rm  }\vec P_i ,{\rm  }\vec R_i $
 and $m,{\rm  }\vec p_{iz}, {\rm  }\vec r_{iz} $
 are mass, momenta and positions of nuclei and electrons respectively and
\begin{equation}
\vec A(\vec R) = \left( {\frac{{2\pi \hbar c^2 }}{{\omega V}}} \right)^{1/2} \sum\limits_{s = 1,2} {\left[ {\vec \varepsilon _s e^{ + i\vec k \cdot \vec R} c_s  + \vec \varepsilon _s^* e^{ - i\vec k \cdot \vec R} c_s^\dag  } \right]} 
\label{eq:2.2}
\end{equation}
is the electromagnetic vector potential in the Coulomb gauge. The potential $ V\left( {\left\{ {\vec R_i } , {\vec r_{iz} } \right\}} \right )$ describes all the Coulomb interactions and can be decomposed in an internal Coulomb interaction $v$, that determines the internal structure of the atoms, and an interatomic potential $V_{ij}$ such that
\begin{equation}
V\left( {\left\{ {\vec R_i } , {\vec r_{iz} } \right\}} \right) = \sum\limits_{i = 1}^N {\left\{ {v\left( {\vec r_{i1}  - \vec R_i ,\vec r_{i2}  - \vec R_i ,...,\vec r_{iZ}  - \vec R_i } \right) + \sum\limits_{j \ne i}^N {V_{ij} } } \right\}} .
\label{eq:2.2bis}
\end{equation}
\end{widetext}	
By means of the unitary transformation \footnote{ Note that Eq. (\ref{subeq:2.3b}) differs from Eq. (5) of  Ref. \onlinecite{bbr} in that we introduce directly the dipole approximation. However this difference is irrelevant for the final result.}
\begin{subequations}
\label{eq:2.3}
	\begin{eqnarray}
\tilde H &=& UHU^{ - 1} \label{subeq:2.3a} \\ 
 U& =& \exp \left[ {\frac{{ie}}{{\hbar c}}\sum\limits_{i = 1}^N {\sum\limits_{z = 1}^Z {\left( {\vec r_{iz}  - \vec R_i } \right) \cdot \vec A\left( {\vec R_i } \right)} } } \right] \label{subeq:2.3b}
\end{eqnarray}
\end{subequations}
and using a result contained in Ref. \onlinecite{wanghioe1}, in Ref. \onlinecite{bbr} it is argued that the partition function can be made independent, in the thermodynamic limit, from the field-matter coupling term at least in the dipole approximation and, consequently, no field condensation can occur. This result, apart from its counterintuitive character \footnote{this conclusion would cast doubts on the validity of the theory of laser.}, is based on the validity of specific convergence properties of operator series (assumptions i and ii, page 832 of Ref. \onlinecite{wanghioe1}) that have not been proven to be valid. In the following we provide a counterexample to the result of Ref. \onlinecite{bbr}.

A correct approach to the consequences of the canonical transformation $UHU^{ - 1} $
 of Ref. \onlinecite{bbr} has been carried out by Widom {\it et al.} \cite{widom}, who explicitly compute the transformed Hamiltonian that we report here for convenience:
\begin{widetext}
	\begin{equation}
\tilde H = \sum\limits_{i = 1}^N {\left\{ {\left[ {\frac{{\vec P_i^2 }}{{2M}} + \sum\limits_{j \ne i}^N {\tilde V_{ij} } } \right] + \left[ \sum\limits_{z = 1}^Z {{\frac{{\vec p_{iz}^2 }}{{2m}} + v\left( {\vec \rho _{i1} ,\vec \rho _{i2,...,} \vec \rho _{iZ} } \right)} }\right]  + \vec \mu _i  \cdot \vec E\left( {\vec R_i } \right)} \right\}}  + H_F ,
\label{eq:2.4}
\end{equation}
\end{widetext}
where $\vec \rho _{iz}  = \vec r_{iz}  - \vec R_{iz} $ are the relative coordinates,
	\begin{equation}
\vec \mu _i  = e\sum\limits_{z = 1}^Z {\vec \rho _{iz} } 
\label{eq:2.5}
\end{equation}
is the electric dipole operator of the $i^{th}$ atom,$
\sum\limits_{i,j \ne i}^N {\tilde V_{ij} } $
 is the screened interatomic potential (i.e. deprived of its dipolar contribution),  $
v\left( {\vec \rho _{i1} ,\vec \rho _{i2,...,} \vec \rho _{iZ} } \right)$
 is the single-atom internal potential \footnote{The term $W_{mol;j}$, Eq. (42) of Ref. \onlinecite{widom} has been neglected because its contribution vanishes when $N$ tends to infinity.} and
	\begin{equation}
\vec E(\vec R) = i\left( {\frac{{2\pi \hbar \rho \omega }}{N}} \right)^{1/2} \sum\limits_{s = 1,2} {\left[ {\vec \varepsilon _s e^{ + i\vec k \cdot \vec R} c_s  - \vec \varepsilon _s^* e^{ - i\vec k \cdot \vec R} c_s^\dag  } \right]} 
\label{eq:2.6}
\end{equation}
is the electric field, where the particle density $\rho  = \frac{N}{V}$
 has been introduced in order to make explicit the dependence on the number of atoms $N$.
We now restrict ourselves to the case where the original interatomic potential is strictly dipolar, i.e. when higher order contributions are negligible  \footnote{ This is a good approximation if the density is below a suitable value, like in the case of BECs.}. This allows us to write:
	\begin{equation}
\tilde V_{ij}  = 0,\ {\rm      }i \ne j.
\label{eq:2.7}
\end{equation}
	We want now to show explicitly that, even in the thermodynamic limit, the dipolar interaction $
\sum\limits_{i = 1}^N {\vec \mu _i \cdot\vec E\left( {\vec R_i } \right)} 
$
 gives a non-negligible contribution to the energy per particle. To this end we explicitly define a superposition of eigenstates of the single-atom Hamiltonian $H_{atom,i}={\frac{{\vec P_i^2 }}{{2M}} } +H_{el,i}$ where
\begin{equation}
H_{el,i}= \sum\limits_{z = 1}^Z {{\frac{{\vec p_{iz}^2 }}{{2m}} + v\left( {\vec \rho _{i1} ,\vec \rho _{i2,...,} \vec \rho _{iZ} } \right)} }  
\label{eq:2.7bis}
\end{equation}
 and compute its energy content. Define the single-atom wave-function
\begin{widetext}
	\begin{equation}
 \psi _\delta  \left( {\vec R,\vec \rho _1 ,\vec \rho _2 ,...,\vec \rho _Z } \right)    
  = i\sin \delta \left\langle {{\vec \rho _1 ,\vec \rho _2 ,...,\vec \rho _Z }}
 \mathrel{\left | {\vphantom {{\vec \rho _1 ,\vec \rho _2 ,...,\vec \rho _Z } {\varphi _p }}}
 \right. \kern-\nulldelimiterspace}
 {{\varphi _p }} \right\rangle \frac{{e^{ - i\frac{{\vec k \cdot \vec R}}{2}} }}{{\sqrt V }} + \cos \delta \left\langle {{\vec \rho _1 ,\vec \rho _2 ,...,\vec \rho _Z }}
 \mathrel{\left | {\vphantom {{\vec \rho _1 ,\vec \rho _2 ,...,\vec \rho _Z } {\varphi _s }}}
 \right. \kern-\nulldelimiterspace}
 {{\varphi _s }} \right\rangle \frac{{e^{i\frac{{\vec k \cdot \vec R}}{2}} }}{{\sqrt V }} \\ 
\label{eq:2.8}
\end{equation}
where $0 \le \delta  \le \frac{\pi }{2}$
 is a free parameter, $\left| {\varphi _s } \right\rangle $
 is the lowest-energy eigenstate and $\left| {\varphi _p } \right\rangle $
 some excited eigenstate of $H_{el}$ for which we have
	\begin{equation}
\mu _0 \vec \varepsilon _1  = \left\langle {\varphi _p } \right|\vec \mu \left| {\varphi _s } \right\rangle  \ne 0,\ \ {\rm    }\mu _0  > 0.
\label{eq:2.9}
\end{equation}
The full quantum state is defined by:
	\begin{equation}
\left| {\Omega \left( {\alpha ,\delta } \right)} \right\rangle  = \left| {\sqrt N \alpha } \right\rangle _{em} \int {\prod\limits_{i = 1}^N {\psi _\delta  \left( {\vec R_i ,\vec \rho _{i1} ,\vec \rho _{i2} ,...,\vec \rho _{iZ} } \right)d^3 \vec R_i \left| {\vec R_i } \right\rangle_i \prod\limits_{z = 1}^Z {d^3 \vec \rho _{iz} \left| {\vec \rho _{iz} } \right\rangle_{iz} } } } 
\label{eq:2.10}
\end{equation}
where $\left| {\sqrt N \alpha } \right\rangle _{em} $
 is a Glauber state \cite{glauber} for which  $c_s \left| {\sqrt N \alpha } \right\rangle _{em}  = \delta _{s1} \sqrt N \alpha \left| {\sqrt N \alpha } \right\rangle _{em} $, $\alpha  > 0.$
 By computing the expectation value of the Hamiltonian $\tilde H$
 on the state $\left| {\Omega \left( {\alpha ,\delta } \right)} \right\rangle $
 we obtain:
	\begin{equation}
\frac{{E(\alpha ,\delta )}}{N} = \frac{1}{N}\left\langle {\Omega \left( {\alpha ,\delta } \right)} \right|\tilde H\left| {\Omega \left( {\alpha ,\delta } \right)} \right\rangle  = \frac{{\hbar ^2 k^2 }}{{8M}} + E_s  + \hbar \omega \left( {\sin ^2 \delta  + \alpha ^2  - \gamma \alpha \sin 2\delta } \right),
\label{eq:2.11}
\end{equation}
\end{widetext}
where $E_s $ and $E_p $ are the single-atom eigenvalues of the states $\left| {\varphi _s } \right\rangle $
 and $\left| {\varphi _p } \right\rangle $
 respectively and we have set $\hbar \omega  = E_p  - E_s $
 and
	\begin{equation}
\gamma  = \mu _0 \sqrt {\frac{{2\pi \rho }}{{\hbar \omega }}} .
\label{eq:2.12}
\end{equation}
We now search for a minimum of $E(\alpha ,\delta )$ with respect to $\alpha $
 and $\delta $. The condition for the minimum is expressed by
\begin{subequations}
\label{eq:2.13}
	\begin{eqnarray}
 \cos (2\delta _{\min })  = \gamma ^{ - 2}  \\ 
 \alpha _{\min }  = \frac{1}{2}\sqrt {\gamma ^2  - \gamma ^{ - 2} } , 
 \end{eqnarray}
\end{subequations}
which has real solutions for $\gamma  \ge 1$. Substitution into eq. (\ref{eq:2.11}) brings to the expression for the minimum energy:
	\begin{equation}
E_{\min }  = \frac{{\hbar ^2 \omega ^2 }}{{8Mc^2 }} + E_s  + \frac{{\hbar \omega }}{4}\left( {2 - \gamma ^2  - \gamma ^{ - 2} } \right)
\label{eq:2.14}
\end{equation}
which is lower than $E_s $
, the 'perturbative' energy, when \footnote{ This result seems to bring to a severe singularity since energy density appears to be unbounded from below. This singularity stems from having neglected the short-range part of the interatomic potential $\tilde V_{ij} $ that results negligible for large average interatomic distances but becomes important when atoms get close enough. As a result there exist a well-defined density corresponding to the minimum energy.}
	\begin{equation}
\gamma ^2  > 1 + \frac{1}{2}\left( {\frac{{\hbar \omega }}{{2Mc^2 }} + \sqrt {\left( {\frac{{\hbar \omega }}{{2Mc^2 }} + 2} \right)^2  - 4} } \right).
\label{eq:2.15}
\end{equation}
In practical cases, being usually $\hbar \omega  <  < Mc^2$, condition  (\ref{eq:2.15}) simplifies to $\gamma ^2  > 1$ or
	\begin{equation}
\rho  > \frac{{\hbar \omega }}{{2\pi \mu _0^2 }},
\label{eq:2.16}
\end{equation}
which represents the condition for the existence of a coherent em field in matter at zero temperature and for the appearance of an {\it energy gap}, whose value depends on the system parameters. Note that such a condition means that the state $\left| {\Omega \left( {\alpha _{\min } ,\delta _{\min } } \right)} \right\rangle $
will be relevant in the thermodynamics of the system for sufficiently high density values only.

In conclusion we have shown that there exist quantum configurations of the field+matter system whose energy content is {\it lower} than that of the lowest energy state of matter uncoupled to the electromagnetic field \footnote{ In Ref. \onlinecite{bbr} this state is expected to be the lowest energy state of the system.} that contribute to the partition function, implying that the conclusions outlined in Ref. \onlinecite{bbr}  cannot be correct, even in the case of a multilevel system. Please note that in the present calculation {\it no use of the result of Ref. \onlinecite{wanghioe1} has been done}.
\section{\label{sec:level12} Conclusions \protect}
We have shown that:
\begin{itemize}
\item the long-standing assertion of  Ref. \onlinecite{bbr} on the impossibility of a superradiant phase transition is not correct,
\item the configurations allowing for  condensation {\it must} have a structure similar to that of eq. (\ref{eq:2.8}). In facts it is essential that, when computing the expectation value with respect to the atomic coordinates $\vec R_i $ and $\vec \rho _{iz} $, the interaction term has a contribution of order $N$, i.e.
	\begin{equation}
\left\langle {\sum\nolimits_{i = 1}^N {\vec \mu _i \cdot\vec E(\vec R_i )} } \right\rangle _{matter}  =  - N\sin (2\delta) \mu _0 \vec \varepsilon _1 \cdot\vec E(\vec 0),
\label{eq:3.1}
\end{equation}
 and this can only happen if the single-particle wave-function that compose the collective state $\left| {\Omega \left( {\alpha ,\delta } \right)} \right\rangle  $ have {\it all} the same phase \footnote{This issue has been recently observed experimentally in Bose-Einstein condensates (BEC), where definite interference patterns have been identified \cite{bose4}, a feature that can only be explained with a phase locking of the constituent atoms \cite{prepbose}.} and the matter wave function has a periodic modulation with period $\frac{{4\pi }}{k}$. In particular the $i^{th}$ atom is in a configuration totally delocalized in space, in contrast with its particle-like character of its incoherent counterpart. However we should emphasize that this result is strictly valid only when $\tilde V_{ij} $ is negligible. Actually, since the system tends to a state where the density is as large as possible, it will reach a density such as to take into play the effect of short-range interactions. As a result, if dipolar coupling $\mu_0$ is sufficiently large, there exist an equilibrium density $\rho^*$ for which the total energy has a minimum,
\item the wave function of the single atoms in the coherent state is different from that of the lowest energy state without em condensate, since it contains a certain fraction of the excited state $\left| {\varphi _p } \right\rangle $.
This very important issue implies that, when matter interacts with external fields, it may exhibit unexpected behavior, a phenomenon referred to as {\it violation of asymptotic freedom } \cite{prepqed}.
\end{itemize}
The inconsistency of the no-go theorem of Ref. \onlinecite{bbr} fully restores theoretical consistency to the work of  Dicke \cite{dicke}, Hepp and Lieb \cite{hepplieb1}, and Preparata \cite{prep1}. 

\begin{acknowledgments}
 We gratefully acknowledge A.Widom for useful discussions on the thermodynamics of the interacting em field and F.Fontana for helpful discussions and for his continuous encouragement in the development of this work. We also are indebted to R.Mele for her careful review of the manuscript and comments. 
\end{acknowledgments}
\newpage 
\bibliography{polacchi}

\begin{thebibliography}{12}
\expandafter\ifx\csname natexlab\endcsname\relax\def\natexlab#1{#1}\fi
\expandafter\ifx\csname bibnamefont\endcsname\relax
  \def\bibnamefont#1{#1}\fi
\expandafter\ifx\csname bibfnamefont\endcsname\relax
  \def\bibfnamefont#1{#1}\fi
\expandafter\ifx\csname citenamefont\endcsname\relax
  \def\citenamefont#1{#1}\fi
\expandafter\ifx\csname url\endcsname\relax
  \def\url#1{\texttt{#1}}\fi
\expandafter\ifx\csname urlprefix\endcsname\relax\def\urlprefix{URL }\fi
\providecommand{\bibinfo}[2]{#2}
\providecommand{\eprint}[2][]{\url{#2}}

\bibitem[{\citenamefont{Dicke}(1954)}]{dicke}
\bibinfo{author}{\bibfnamefont{R.~H.} \bibnamefont{Dicke}}, \bibinfo{journal}{Phys.\ Rev.} \textbf{\bibinfo{volume}{93}}, \bibinfo{pages}{99} (\bibinfo{year}{1954}).

\bibitem[{\citenamefont{Hepp and Lieb}(1973)}]{hepplieb1}
\bibinfo{author}{\bibfnamefont{K.}~\bibnamefont{Hepp}} \bibnamefont{and} \bibinfo{author}{\bibfnamefont{E.~H.} \bibnamefont{Lieb}}, \bibinfo{journal}{Ann.\ Phys.} \textbf{\bibinfo{volume}{76}}, \bibinfo{pages}{360} (\bibinfo{year}{1973}).

\bibitem[{\citenamefont{Wang and Hioe}(1973)}]{wanghioe1}
\bibinfo{author}{\bibfnamefont{Y.~K.} \bibnamefont{Wang}} \bibnamefont{and} \bibinfo{author}{\bibfnamefont{F.~T.} \bibnamefont{Hioe}}, \bibinfo{journal}{Phys. Rev. A} \textbf{\bibinfo{volume}{7}}, \bibinfo{pages}{831} (\bibinfo{year}{1973}).

\bibitem[{\citenamefont{Preparata}(1995)}]{prep1}
\bibinfo{author}{\bibfnamefont{G.}~\bibnamefont{Preparata}}, \emph{\bibinfo{title}{QED Coherence in Matter}} (\bibinfo{publisher}{World Scientific}, \bibinfo{year}{1995}).

\bibitem[{\citenamefont{Bialynicki-Birula and K.Rzazewski}(1979)}]{bbr}
\bibinfo{author}{\bibfnamefont{I.}~\bibnamefont{Bialynicki-Birula}} \bibnamefont{and} \bibinfo{author}{\bibnamefont{K.Rzazewski}}, \bibinfo{journal}{Phys. Rev. A} \textbf{\bibinfo{volume}{19}}, \bibinfo{pages}{301} (\bibinfo{year}{1979}).

\bibitem[{\citenamefont{Giudice et~al.}(1993)\citenamefont{Giudice, Mele, and Preparata}}]{prepmele}
\bibinfo{author}{\bibfnamefont{E.~D.} \bibnamefont{Giudice}}, \bibinfo{author}{\bibfnamefont{R.}~\bibnamefont{Mele}}, \bibnamefont{and} \bibinfo{author}{\bibfnamefont{G.}~\bibnamefont{Preparata}}, \bibinfo{journal}{Mod. Phys. Lett B} \textbf{\bibinfo{volume}{7}}, \bibinfo{pages}{1851} (\bibinfo{year}{1993}).

\bibitem[{\citenamefont{Enz}(1997)}]{enz}
\bibinfo{author}{\bibfnamefont{C.~P.} \bibnamefont{Enz}}, \bibinfo{journal}{Helv. Phys. Acta} \textbf{\bibinfo{volume}{70}}, \bibinfo{pages}{141} (\bibinfo{year}{1997}).

\bibitem[{\citenamefont{Sivasubramanian et~al.}(2001)\citenamefont{Sivasubramanian, Widom, and Srivastava}}]{widom}
\bibinfo{author}{\bibfnamefont{S.}~\bibnamefont{Sivasubramanian}}, \bibinfo{author}{\bibfnamefont{A.}~\bibnamefont{Widom}}, \bibnamefont{and} \bibinfo{author}{\bibfnamefont{Y.~N.} \bibnamefont{Srivastava}}, \bibinfo{journal}{Physica A} \textbf{\bibinfo{volume}{301}}, \bibinfo{pages}{241} (\bibinfo{year}{2001}).

\bibitem[{\citenamefont{Glauber}(1963)}]{glauber}
\bibinfo{author}{\bibfnamefont{R.}~\bibnamefont{Glauber}}, \bibinfo{journal}{Phys. Rev.} \textbf{\bibinfo{volume}{131}}, \bibinfo{pages}{2766} (\bibinfo{year}{1963}).

\bibitem[{\citenamefont{Preparata}(1990)}]{prepqed}
\bibinfo{author}{\bibfnamefont{G.}~\bibnamefont{Preparata}}, in \emph{\bibinfo{booktitle}{Problems of Fundamental Modern Physics}} (\bibinfo{publisher}{R. Cherubini, P. Dalpiaz and B. Minetti, World Scientific, Singapore}, \bibinfo{year}{1990}).

\bibitem[{\citenamefont{M.R.Andrews et~al.}(1997)\citenamefont{M.R.Andrews, Townsend, Miesner, Durfee, Kurn, and Ketterle}}]{bose4}
\bibinfo{author}{\bibnamefont{M.R.Andrews}}, \bibinfo{author}{\bibfnamefont{C.}~\bibnamefont{Townsend}}, \bibinfo{author}{\bibfnamefont{H.}~\bibnamefont{Miesner}}, \bibinfo{author}{\bibfnamefont{D.}~\bibnamefont{Durfee}}, \bibinfo{author}{\bibfnamefont{D.}~\bibnamefont{Kurn}}, \bibnamefont{and} \bibinfo{author}{\bibfnamefont{W.}~\bibnamefont{Ketterle}}, \bibinfo{journal}{Science} \textbf{\bibinfo{volume}{275}}, \bibinfo{pages}{589} (\bibinfo{year}{1997}).

\bibitem[{\citenamefont{Giudice and Preparata}()}]{prepbose}
\bibinfo{author}{\bibfnamefont{E.~D.} \bibnamefont{Giudice}} \bibnamefont{and} \bibinfo{author}{\bibfnamefont{G.}~\bibnamefont{Preparata}}, \eprint{cond-mat/9812018}.

\end{thebibliography}

\end{document}